\documentclass[manuscript]{acmart}

\AtBeginDocument{%
  }

\setcopyright{acmlicensed}
\copyrightyear{2026}
\acmYear{2026}
\acmDOI{XXXXXXX.XXXXXXX}

\acmConference[CHI '26]{Proceedings of the CHI Conference on Human Factors in Computing Systems}{April 13--17, 2026}{Barcelona, Spain}
\acmBooktitle{Proceedings of the CHI Conference on Human Factors in Computing Systems (CHI '26), April 13--17, 2026, Barcelona, Spain}

\acmISBN{978-1-4503-XXXX-X/2018/06}

\usepackage{csquotes}
\usepackage{todonotes}
\usepackage{soul}
\usepackage{multirow}
\usepackage{booktabs}
\usepackage{array}
\usepackage{colortbl}
\usepackage{makecell}
\usepackage{subcaption}
\usepackage{colortbl}
\usepackage[table]{xcolor}
\usepackage{array}
\usepackage{multirow}
\usepackage{caption}
\begin{document}

% \title{Exploring the Use of LLMs for Synthetic Cognitive Walkthrough}
\title{Synthetic Cognitive Walkthrough: Aligning Large Language Model's Performance with Human Cognitive Walkthrough}
\author{Ruican Zhong}
\orcid{0009-0004-7169-0675}
\email{rzhong98@uw.edu}
\affiliation{%
  \institution{University of Washington}
  \city{Seattle}
  \state{WA}
  \country{USA}}
\author{David W. McDonald}
\orcid{0000-0001-5882-828X}
\email{dwmcphd@gmail.com}
\affiliation{%
  \institution{University of Washington}
  \city{Seattle}
  \state{WA}
  \country{USA}}

\author{Gary Hsieh}
\orcid{0000-0002-9460-2568}
\email{garyhs@uw.edu}
\affiliation{%
  \institution{University of Washington}
  \city{Seattle}
  \state{WA}
  \country{USA}}

% Synthetic Cognitive Walkthrough: Exploring the Alignment
% Aligning the LLM's Performance with Human Cognitive Walkthrough

\begin{abstract}
%Conducting usability testing, such as cognitive walkthrough (CW), can be costly. Recent developments in large language models (LLMs), with capabilities in visual reasoning and UI navigation, present an opportunity to automate CW. Our work explores whether LLMs can simulate human behavior in conducting CW. We found that while the LLM can navigate the interface and provide reasonable rationales for its decisions, its behavior was fundamentally different from that of humans. LLMs were more capable of completing the tasks compared to human users (100\% to 88.3\%). Their navigation paths were more aligned with the correct paths than those of humans. And they were not able to identify as many usability issues as humans did. In a follow-up study, we explored prompting strategies to explicitly guide LLMs in detecting usability issues and demonstrated their effectiveness in predicting human-identified confusion points. This shows that with more careful design of synthetic CW methods, LLMs can potentially be leveraged to scale usability evaluation and help address resource constraints, despite that it cannot replicate human behavior.
Conducting usability testing like cognitive walkthrough (CW) can be costly. Recent developments in large language models (LLMs), with visual reasoning and UI navigation capabilities, present opportunities to automate CW. We explored whether LLMs (GPT-4 and Gemini-2.5-pro) can simulate human behavior in CW by comparing their walkthroughs with human participants. While LLMs could navigate interfaces and provide reasonable rationales, their behavior differed from humans. LLM-prompted CW achieved higher task completion rates than humans and followed more optimal navigation paths, while identifying fewer potential failure points. However, follow-up studies demonstrated that with additional prompting, LLMs can predict human-identified failure points, aligning their performance with human participants. Our work highlights that while LLMs may not replicate human behaviors exactly, they can be leveraged for scaling usability walkthroughs and providing UI insights, offering a valuable complement to traditional usability testing.

\end{abstract}

\maketitle
\section{Introduction}
The advent of Large Language Models (LLMs) with advanced reasoning capabilities has opened new possibilities for automated user interface (UI) evaluation. Instead of conducting user evaluation studies with people, LLMs, having been trained on a large corpus of human-generated data, may be able to provide sufficiently good approximations of human usage and feedback to obviate the need for human involvement in identifying potential interface issues. This is a potentially transformative change to usability testing. Despite the many benefits of user evaluation and its prevalence in industry, recruiting, scheduling, working with, and compensating participants often require significant time and resources. If UI evaluation can be automated, this could reduce cost, enable more agile design and evaluation processes, and allow for more design iterations.  

A growing body of research has been published demonstrating that LLM-powered agents, especially those with more UI training data, can have a sufficiently good understanding of the consequences of their UI actions ~\cite{li2022spotlight,jiang2025iluvui,you2024ferret}. However, what remains unclear is whether this can be effectively translated to support UI evaluation. Just because an agent can plan specific actions and navigate through interfaces does not mean it is doing so in the same way a human user would. Potential differences may matter in the user evaluation context, since the goal is not to have the AI identify optimal paths through the UI and perform tasks, but rather to simulate human experiences so that designers can identify potential usability issues that a user may experience. What is missing, therefore, are rigorous comparative studies exploring whether and how LLMs' navigation of interfaces differs from humans'.

To address this gap, we explore the use of LLMs for cognitive walkthroughs (CW). CW is a commonly used usability method to help evaluate a system's learnability and ease of use ~\cite{mahatody2010state}. Unlike other usability inspection methods, such as heuristic evaluation that uses heuristics to identify general interface problems, CW is specifically focused on how users navigate the system and accomplish tasks ~\cite{lewis1997cognitive}. Thus, to demonstrate the efficacy of using LLMs to automate CW, we must first show that the LLMs' action sequences, or navigation steps, are comparable to those of human evaluators. This would then allow user researchers to use LLMs to assess the learnability of the interface: whether potential human users would notice that the correct action is accessible on a given screen, and whether the users would understand that the desired subtask outcome can be achieved by the action, etc. 

We begin by iterating on prompts and workflows for using LLMs for CW. Specifically, we crafted the prompt so that the LLM could reasonably navigate the interface and articulate the rationale behind its actions. We then conducted a comparative study contrasting the CW results between LLMs (GPT-4 and Gemini-2.5-pro) and 10 human evaluators on two mobile apps on these key dimensions of CW: task completion rate, path navigation alignment, and number of potential failure points. We found that LLM-prompted CW achieved higher task completion rates (100\% for GPT-4 and 97.2\% for Gemini-2.5-pro) than humans (88.2\%). LLMs took fewer steps to achieve the tasks, and their paths were more aligned with the correct paths. Further, all humans identified 10 potential failure points, but all LLM runs aggregatedly identified 3. This demonstrates that the LLM outperformed humans in navigating the apps, but did not simulate humans' behavior. Analyses of navigation paths and think-aloud further suggest the following behavioral differences between humans and LLM: 1) human evaluators were more likely to take a breadth-first search when they were uncertain about the next steps, 2) humans were more prone to making mistakes due to memory issues. 

% Thus, our comparative study shows that the LLM is not conducting CW similar to how humans would do them. 
% We explored whether LLMs (GPT-4 and Gemini-2.5-pro) can simulate human behavior in CW by comparing their walkthroughs with human participants. While LLMs could navigate interfaces and provide reasonable rationales, their behavior differed from humans. 

% In terms of path navigation, we found the 

To further explore whether the LLM can be prompted to predict the potential navigational failure points (given that is one of the key objectives of CW), we conducted a follow-up study, exploring two approaches. One is to explicitly prompt the LLM to identify potential failure points as it navigates through the interface during CW. Another approach, dubbed the without context approach, is where we simply ask the LLMs the same question, but only with individual screens, without having it perform walkthroughs. We found that the without-context approach was less effective and less consistent at predicting the human-identified potential failure points. In contrast, the with-context approach was able to consistently predict these failure points across our runs. 
% In fact, the LLM was able to outperform human in task completion rate and its paths were more aligned with the correct path. 
% and to use that rating as an indicator for the failure points reported by humans
Overall, our work reveals that the LLM does not conduct CW in a similar way to humans. However, as LLMs walk through the screens, they can be prompted to identify potential failure points that humans would have identified and offer think-aloud insights on interface design issues, suggesting an opportunity to use LLMs to support the objectives of CW. 

Our research makes the following contributions:
\begin{itemize}
    \item Empirical demonstration that current reasoning models can correctly infer action steps and perform tasks, at a level that is comparable to or even better than human users.  
    \item Comparative qualitative insights into how LLMs and humans differ in conducting UI walkthroughs.
    \item An LLM-powered approach that can help identify potential failure points in CW.
    % \item An LLM-powered approach that can help identify task-specific user experience problems. 
\end{itemize}
\section{Related Work}
\label{section:related-work}
Here, we outline two areas of prior work that we build on: 1) cognitive walkthrough and automated user evaluation, 2) UI understanding, navigation, and user simulations. 

\subsection{Cognitive Walkthrough and Automated User Evaluation}

Cognitive walkthrough (CW) is a commonly used usability method, focused on exploring the overall user journey through the interface and assessing whether a new user can easily accomplish the intended tasks ~\cite{mahatody2010state}. CW evaluators (either real users or other domain experts) are asked to ``walkthrough'' the interface to accomplish tasks. Evaluators are encouraged to think-aloud to make their actions visible through words. Throughout each step, evaluators' behaviors are used to answer four key questions ~\cite{flaherty2022how}: (1) Will users try to achieve the right result?; (2) Will users notice that the correct action is available?; (3) Will users associate the correct action with the result they’re trying to achieve?; and (4) After the action is performed, will users see that progress is made toward the goal? With the data collected, researchers can uncover the following dimensions and ultimately understand how users conceptualize the design: 1) task completion success rate, 2) whether the users' paths taken are aligned with the correct paths, and 3) the potential failure points that reflect potential usability issues. Among the three dimensions, the potential failure points contribute significantly to the evaluation result, as they directly point to potential breakdowns in the learnability of UIs, which is the key purpose of CW. As a formative evaluation strategy, CW can be conducted with early prototypes such as screenshots of the interface, using a wizard-of-oz approach, where the facilitator presents the appropriate screenshots given evaluators' interactions.  
% identify potential failure points, discoverability issues, and 

% 1. What is cognitive walkthrough, what do designers gain from CW? What are the key metrics/outcomes from CW?

As CW and other usability methods become increasingly utilized in human-centered design processes, many have explored strategies and tools to reduce the time and human resources needed to perform these usability methods. In their 2001 review of the then state-of-the-art automated usability methods ~\cite{ivory2001state}, Ivory and Hearst found that automation in general is greatly underexplored. Of about 110 usability evaluation methods they surveyed, only 33\% had some sort of automation support. Within those, the level of automated support is also limited. For example, specific to CW, the only tool that had been developed was a software to simply assist with the documentation of the walkthrough ~\cite{rieman1991automated}, and it still requires formal interface use --- far from an automated CW that can significantly reduce evaluation costs. 

One inherent challenge with fully automating the evaluator's walkthrough of an interface is that the automated tool must have a good understanding of a given interface and be able to approximate human behaviors. This resulted in earlier work on developing formal grammar to represent the UI design, and the use of Model Human Processor to simulate human behaviors to predict task performance ~\cite{john1996goms, payne1986task}. However, for interfaces with complex interactions, these models are time-intensive to build and maintain, and have resulted in limited uptake in these approaches.

\subsection{UI Understanding, Navigation, and User Simulation}

The rise of machine learning has contributed to significant advances in overcoming the problem of UI understanding. Using various deep learning techniques and different annotated datasets, researchers have demonstrated the feasibility of determining user interface elements and their interactivity for mobile and web interfaces ~\cite{deka2017rico,wang2021screen2words,li2021screen2vec,li2020mapping, swearngin2019modeling}. In addition to approaches that rely on structural and spatial data from the interface directly for UI understanding, research has also explored the use of only pixel-based information ~\cite{zhang2021screen,bunian2021vins,sunkara2022towards,li2022learning}, paving the way for a more generalizable approach for automated UI understanding using images of the interface directly. More recently, LLMs and multimodal LLMs (MLLMs) have further advanced UI understanding tasks. Research on MLLMs specifically trained for user interfaces shows that they can identify the existence of UI elements, determine UI element types, and assess UI element purpose ~\cite{li2022spotlight,jiang2025iluvui,you2024ferret}. 

Concurrently, researchers have also been examining the problem of UI navigation, particularly for the goal of task automation. Using deep learning, researchers have developed corpus pairing natural language commands to web ~\cite{pasupat2018mapping} and mobile UI action sequences ~\cite{li2020mapping}. Li et al. ~\cite{li2020mapping} showed that their models can achieved 70.59\%
accuracy on predicting ground-truth action sequences. Subsequent research showed that using few-shot prompting with GPT-3 is able to achieve 45\%
accuracy  ~\cite{wang2023enabling}. ResponsibleTA uses LLM-based coordinators and executors and is able to verify the completeness of the commands with an accuracy of 84\%, outperforming GPT-4 (63\%) and ChatGPT/GPT-3.5 (61\%) ~\cite{zhang2023responsible}. AutoDroid also showed that using commonsense knowledge of LLMs and domain-specific knowledge of apps, it is able to generate
actions with an accuracy of 90.9\%, and complete tasks with
a success rate of 71.3\%. Further, they showed that their system outperforms the GPT-4-powered
baselines by 36.4\% and 39.7\% ~\cite{wen2024autodroid}. 

This prior research has provided the foundation for recent applications in using AI-powered agents for usability testing and evaluation. This includes work showing that agents can be used to test the GUI ~\cite{eskonen2020automating}, and inspect UI to examine heuristic violations ~\cite{zhong2025synthetic, guerino2025can, duan2024generating}. 
Most related to our work, research has also explored the use of AI to simulate user flow for user evaluation purposes. This includes SimUser, which uses an LLM-powered user agent to navigate the interface, simulating users' thoughts and reasons along the process. They found that 20 rounds of SimUser feedback covered 70\% of the usage scenarios explored by 48 human participants, and identified 80\% of the usability issues uncovered by humans ~\cite{xiang2024simuser}. UXAgent also uses LLM to simulate Usability Testing of web pages ~\cite{lu2025uxagent}. It does so by generating thousands of simulated users (with different personas) to test the website. UXAgent provides action and reasoning traces to the researchers to help them assess potential usability breakdowns. While participants appreciated the potential of a tool like UXAgent to support early feedback collection for iteration of the study design, participants did find that the reasoning traces produced are hard to read and interpret. Finally, research has also examined the use of GPT-4 to conduct walkthroughs of an "upload a photo" task, and showed its ability to determine appropriate subtasks and actions to complete the tasks, and provide feedback about usability ~\cite{bisante2024enhancing}. 

In summary, these prior works suggest that we now have the technology to automatically understand UIs and to automate UI tasks. However, there are several important unanswered research questions if the goal is to use it to automate CW. First, prior prediction models focused on maximizing accuracy in task completion. However, in the context of CW, where the goal is to evaluate interface learnability, what we need is for LLMs to have comparable task completion rates as humans. If the agents are too unsuccessful (or even too successful) at navigating the interface, they might not help uncover the same set of issues that human users may face. Second, the evaluation focused on task completion also overlooks navigation path differences between AI agents and humans. Finally, practitioners are unlikely to train or fine-tune their own LLM or MLLM. With the advances in frontier reasoning models, there is insufficient empirical data on just how well current off-the-shelf models can perform these evaluation tasks. Thus, in this work, we conducted a comparative study comparing LLMs against human participants to answer these important research questions and to advance our understanding of LLMs for UI evaluation.

% Explain how this related work incorporated the users into CW process compared to the traditional method, which is what we are using in our work. [https://link.springer.com/chapter/10.1007/1-4020-4205-1_20]

% 1. What is cognitive walkthrough, what do designers gain from CW? What are the key metrics/outcomes from CW?
% 2. LLM 
% 1) Use of LLM in usability testing (related work section from the HE paper) [They did not specifically focus on cognitive walkthrough, which is more involved "cognitively" with making a decision about paths, etc.]
% 2) LLM for automatic navigation of mobile interfaces (bring in work about automatic UI testing/navigation techniques) [It's more about asking the LLM to navigate the interface successfully, not to identify usability issues]

% Contribution of our work:
% 1) Identify behavior differences between LLM and human in doing CW
% 2) Showcase how to use LLM driven CW to predict human conducted CW sessions
% 
% consistency test

\section{Cognitive Walkthrough Prompt Design}
\label{section:prompt-design}
% \todo{build out a visual for the system design, what information needs into which components}
To automate CW, we designed two AI agents, one performing the role of the facilitator and one performing the role of the evaluator (participant), mimicking the roles typical of CW sessions. In addition to leading the session, the facilitator-agent also helps select the appropriate screenshots to present to evaluators depending on their actions. The evaluator performs the role of the user and performs the think-aloud as it navigates through the interface.

%we developed a Wizard-of-Oz (WOZ) pipeline. The system takes the following input: a set of screenshots of an app and the user tasks for the CW. Our pipeline has two AI agents (both supported by LLM), one performing the role of the facilitator, and one performing the role of the evaluator (participant), mimicking the roles in in-person sessions. The facilitator leads the session, but also acts as the wizard in our Wizard of Oz design -- selecting the appropriate screenshots to present to evaluators depending on their actions. The evaluator performs the role of the user and performs the think-aloud as they navigate through the interface.

%In our design, the evaluator does not interact with the interface directly but would describe the next action and the facilitator would provide the updated screen. We chose this approach primarily because while there are some research and commercial tools that may allow for our evaluator-agent to navigate  interfaces directly, it is unclear how well they work or how exactly they are built. It would an additional layer in our comparator study between LLMs and human navigation. Thus, we opted to use the existing frontier models directly. %Our facilitator-as-wizard approach does provide the added benefit of enabling the evaluation of early prototypes (e.g., screenshots). 

In this section, we describe how we iteratively designed the prompts for each component of the pipeline and provide the finalized prompts.
% As it is possible to upload screenshots to LLMs but not directly link LLMs to navigate entire applications. 

\subsection{Pipeline Setup}
\subsubsection{Facilitator}
\label{section:facilitator-prompt}
The facilitator acts as a researcher, guiding the evaluator through the CW session. This ensures the evaluator provides formatted output and receives instructions to address specific issues. It provides guidance when the evaluator is stuck in one place, conducts repeated action, or feels uncertain about next steps. 

To design the prompt for this component, we first attempted a naive version where we instructed the LLM to \textit{``act as a facilitator in a CW and provide guidance to users as needed.''} However, this was not effective because it just provided simple instructions such as \textit{``I'm the facilitator of this session. Let's start with this screenshot and let me know what you think.''} When the evaluator did not provide enough explanations, the facilitator was not able to provide additional guidance to elicit more insight. Therefore, we crafted a more detailed prompt, specifying that when the evaluator responses are not clear, the facilitator should \textit{``prompt them to think further, ask them `why and how questions.'''} We then encountered an issue where the facilitator again could not provide enough guidance when the evaluator performed looping behavior (e.g., the evaluator continually goes from screen A -> B -> C -> B -> C, etc.). To address this, we implemented a fail-safe mechanism, which is similar to how a human facilitator would handle this situation. Essentially, when a loop is detected, our facilitator automatically sends the following message to the evaluator: \textit{``The action you provided/identified is not available on the screen. Consider trying a different action here. Please revise your action.''} This instruction was able to help the responder break out of the loop and made the LLM facilitator better simulate how a human facilitator would guide a CW session. The finalized prompt for the facilitator is shown in \autoref{appendix:failitator-prompt}.

In addition, the facilitator also presents the appropriate next screen to the evaluator as it navigates through the interface with the think-aloud. When the LLM evaluator articulates that it wants to click on a component to navigate to the next screen, the facilitator helps pull out that next screen accordingly from a pre-compiled image database (containing all the screens and transitions between the screens) for the tested app, and uses it to move along the CW session. 
% Then, the facilitator will provide that screen to the evaluator and ask for the next steps again. 

% There is an image selector helper that helps search within our JSON dataset and find the corresponding next screen. Essentially, the image selector helper receives the current screen code name and extracts all the possible next actions and next screens from the current screen from the JSON file. Then it uses the following prompt to find the best matching next action and next screen, which then allows the evaluator to navigate the interface:

% We used GPT-4 by OpenAI as the LLM for our pipeline because it is one of the widely used state-of-the-art models, and prior works have demonstrated its ability to navigate mobile UIs and provide design feedback. [CITE] Additionally, to ensure our approach achieved reasonable performance and maintained performance consistency, we iterated our prompts with the LLM multiple times. 
\subsubsection{Evaluator}
The evaluator plays the role of the potential interface user in a CW session. It is given screens of an app and a user task, and asked to analyze and determine where it would like to interact with next to complete the user task and provide a rationale for it. 

To design the prompt for this component, we again started with a basic version where we asked the LLM to \textit{``describe where you would like to interact next to complete the task, and provide the rationale for that action.''} But the evaluator outputs indicate that its path navigation patterns were generally random and without a concrete explanation of why the specific choice would help achieve the task. For example, when prompted to \textit{find a lesson to order food and drink} in a language learning app (\autoref{fig:app1-home}), the evaluator clicked on the \textit{unlock all German courses,} which was not directly related to this task and actually was an advertisement. Additionally, when prompted the LLM to run this task multiple times, the LLM sometimes chose the \textit{profile} icon, and sometimes navigated to \textit{review}, neither of which was closely relevant to the user task. Therefore, a naive version of the prompt could not ensure that the evaluator would reliably navigate the apps. To address this, we added instructions in the evaluator prompt that it should first identify all the plausible next steps from the current screen that could help achieve the task, then rate the likelihood of each option. Then, the evaluator would make a decision based on this analysis. With this mechanism of instructing the LLM to think through all possibilities before making a decision, we observed a great improvement in performance. But then we noticed another issue. When the evaluator gets confused about a specific screen, it gets stuck there or starts looping in the same place. To address this, in addition to having the facilitator provide more guidance (as specified above with the fail-safe), we also added instructions in the evaluator prompt, noting that if it is seeing the same sets of screens, it means that it is looping and should consider other options. With this modification, we observed stable and reasonable performance by the evaluator as it navigates the apps and provides CW-related insights. The finalized prompt can be found in \autoref{appendix:evaluator-prompt}.

% To address this issue, we added in both the facilitator prompt and the evaluator prompt that when the evaluator is stuck on a specific screen, it should think about breaking out of the loop by considering other available options. We also added the failed-safe mechanism mentioned in 

% \subsection{Pipeline Setup}

%\subsubsection{Components}
%There are two components in the pipeline: a facilitator, a evaluator (the user). Under the evaluator, there is an image selector helper that assist the evaluator. Each component is supported by LLM individually. The facilitator and evaluator communicates with each other in using a conversation approach. 

\subsubsection{Pipeline Process}
The CW begins with the user inputting the following to our system: the description of the user task, a start screen for that task, and a database containing all images in the tested app, along with the navigation logic. Using this information, the facilitator presents the task and the start screen to the evaluator, providing guidance about how to conduct a CW. The evaluator responds to the facilitator with how it wants to navigate the screen to achieve the task and why. Then, the facilitator selects the next screen based on the evaluator's response and asks the evaluator to continue with the CW. This process repeats until the task is completed, or if the facilitator identifies that the evaluator is stuck (i.e., in a loop) for more than 5 times. 

\section{Study 1: Comparing LLM and Human Conducted Cognitive Walkthrough}
\begin{figure}[!tbp]
  \centering
  \begin{minipage}[b]{0.47\textwidth}
    \vspace{2pt}
        \centering
        \begin{subfigure}[t]{0.5\textwidth}
            \centering
            \includegraphics[width=\linewidth]{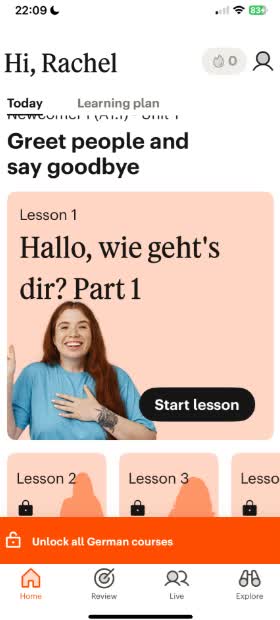}
            \caption{Key components: \textit{learning plan, live, review, explore, profile.}}
            \label{fig:app1-home}
        \end{subfigure}%
        ~ 
        \begin{subfigure}[t]{0.5\textwidth}
            \centering
            \includegraphics[width=\linewidth]{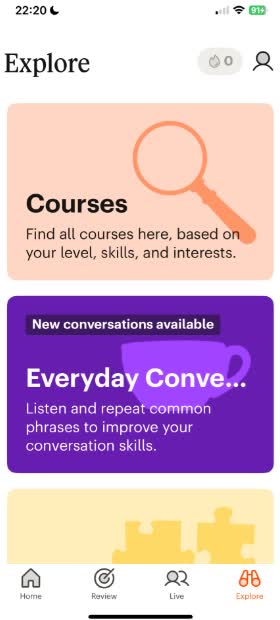}
            \caption{Key components: \textit{courses module} and \textit{everyday conversation module.}}
            \label{fig:app1-explore}
        \end{subfigure}
        \caption{These two figures are example screens in the language learning app.}

    % \label{fig:example1}
  \label{fig:rental}
  \end{minipage}
  \hfill
  \begin{minipage}[b]{0.47\textwidth}
    \vspace{2pt}
        \begin{subfigure}[t]{0.5\textwidth}
            \centering
            \includegraphics[width=\linewidth]{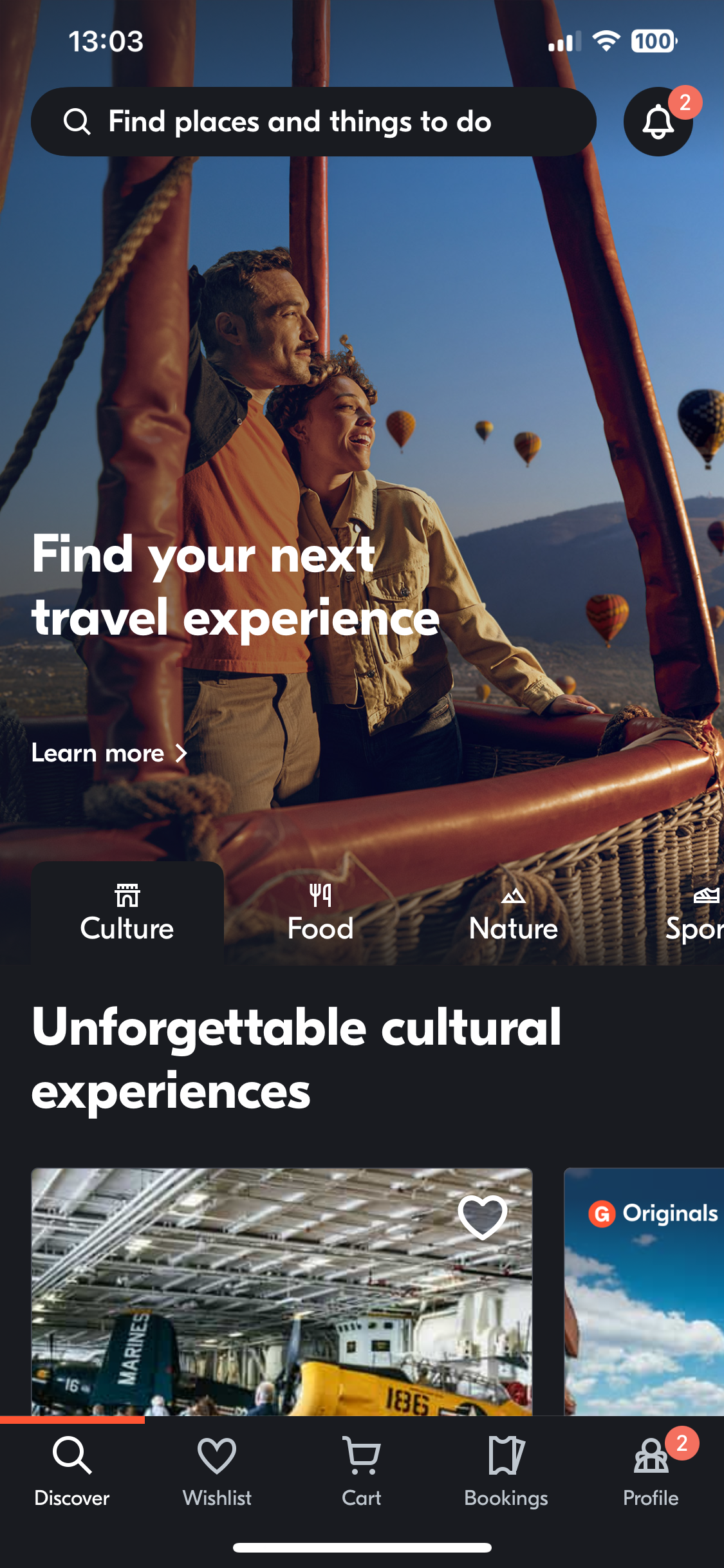}
            \caption{Key components: \textit{search bar} and \textit{culture tab.}}
            \label{fig:app2-home}
        \end{subfigure}%
        ~ 
        \begin{subfigure}[t]{0.5\textwidth}
            \centering
            \includegraphics[width=\linewidth]{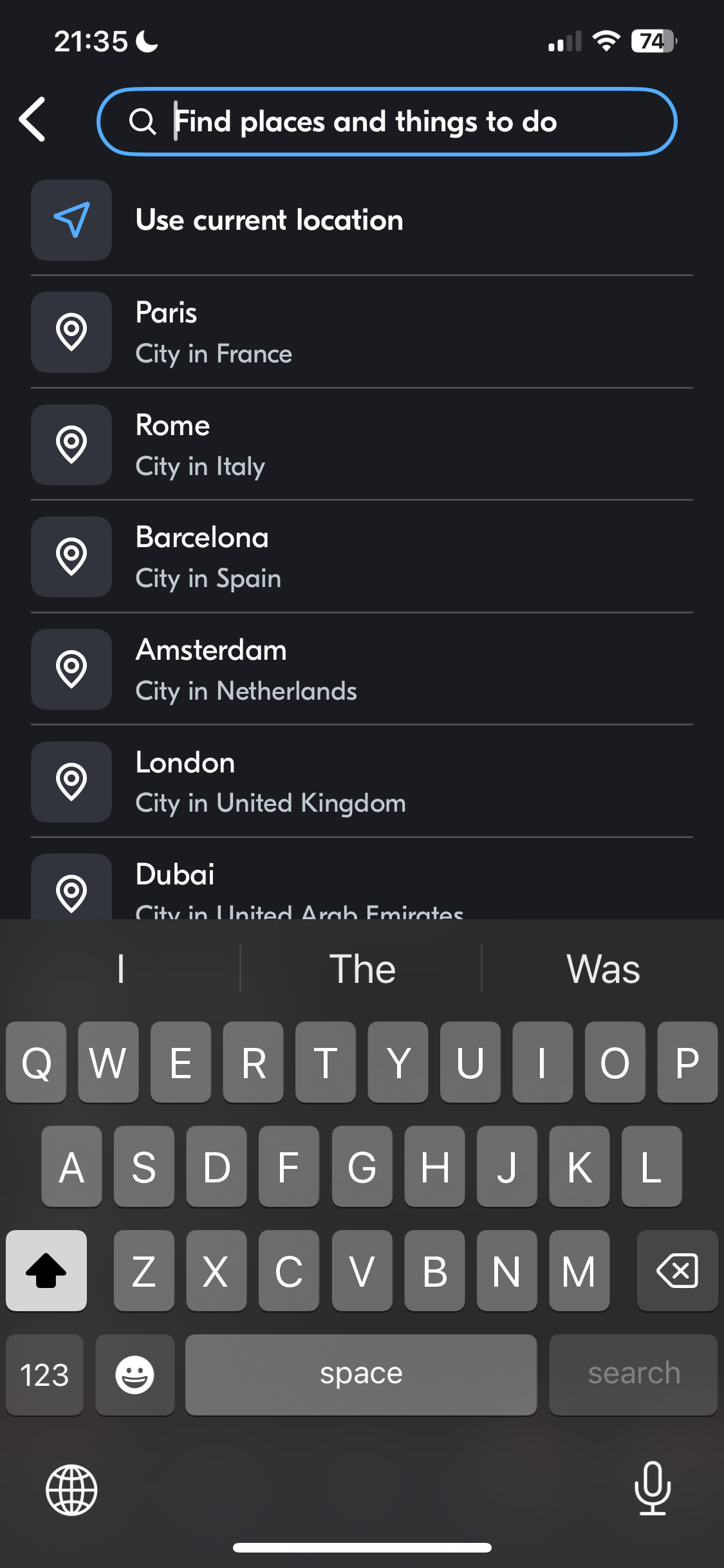}
            \caption{Key components: \textit{search bar} and \textit{Amsterdam} in the location list.}
            \label{fig:app2-search}
        \end{subfigure}
        \label{fig:language}
        \caption{These two figures are example screens in the booking app.}
    \label{fig:booking}
  \end{minipage}
\end{figure}
In this first study, we tested and compared outcomes between the LLM-powered walkthroughs against walkthroughs by human participants. 

%especially about researchers' desired control over the workflow

%As described in \autoref{section:related-work}, there are three key outcomes of a CW session: 1) task completion rate 2) path navigation differences 3) potential failure points identification. 

% \subsubsection{Dataset}

\subsection{Procedure}
To understand the performance of LLMs in conducting CW, we selected two mobile applications: a language learning app and a trip booking app. These two apps are representative of two commonly used categories of mobile apps: Lifestyle and Education~\cite{appcategory}. For each app, we curated a total of 6 user tasks for the CW session, varying in difficulty and complexity. The number of steps required to complete each task ranges from 1 to 38, with an average of 7.4 across the two apps. To design these tasks, the authors closely investigated the apps to identify potential failure points and then designed user tasks around them. This helps ensure that users will encounter these potential failure points when completing the tasks, which will help us collect a robust and diverse dataset. During this process, the authors also curated a correct path set, containing all possible optimal pathways that users could take to accomplish the tasks. This will be used as the reference to compare humans' and LLMs' performances. 

In our study, we generated two sets of evaluation results of the aforementioned user tasks for both apps: an LLM set and a user evaluator set. The LLM and the users had access to the same apps and completed the same user tasks. Our study protocol was approved by the IRB. 

\subsubsection{Synthetic Cognitive Walkthrough (LLM Set)}
Using the pipeline shared in Section~\ref{section:prompt-design}, we asked the LLM to complete the user tasks for both apps. We tested our approach with both GPT-4 by OpenAI~\cite{achiam2023gpt} and Gemini-2.5-pro by Google~\cite{comanici2025gemini}, which are two commonly used state-of-the-art models. In total, for each of the two apps, we asked GPT-4 to evaluate 5 times, and asked Gemini-2.5-pro to evaluate 3 times. This resulted in a total of 16 runs with the LLMs. Next, we refer to them as GPT runs and Gemini runs.

\subsubsection{Human Set}
We recruited our participants on social media to complete the CW of the two apps. No prior experience with CW was required. During the session, participants first completed a survey about their general demographics and prior experience with the type of app they were testing.

In total, we recruited 10 participants; 5 were randomly assigned to evaluate the language learning app, and 5 were assigned to the booking app. The majority (8/10) of the participants were in the age range of 18-23. 4 participants identified as men, 5 as women, and one as gender nonconforming. Most of the participants (8/10) reported their frequency of use of language learning apps and booking apps was at least 2-5 times a month, indicating some experience with similar apps. None of the participants had interacted with the tested apps before.  

During the study, the facilitator first explains the goal of CW and outlines the general process. Then, the participants were given the user tasks for one of the two apps. We provided the participants with the same instructions as those used to prompt the LLM. The following is an example instruction for a task in the booking app:
\begin{quote}
\begin{displayquote}
    \small
    \texttt{Given the following mobile app and the user task (\textit{You would like to find an outdoor day trip in Amsterdam and save it to view later}), please conduct a CW. Describe what you would do, which component you would interact with to achieve this task. Please use the ``think-aloud'' protocol to describe your thought process. Please describe if the current screen contains anything that may confuse you or prevent you from completing the task. Describe why it is confusing.}
\end{displayquote}
\end{quote}

The participant was asked to complete the user task by thinking out loud. The facilitator screenshares the app. And based on the participants' articulation of where they would like to click on/interact with next, the researcher carried out the actions accordingly. Participants also articulated their thought processes and rationales throughout the session. If their explanation was unclear or if it did not fully address the purpose of a CW, the researcher would prompt them to provide a more detailed explanation. 

On average, each session took 55 minutes. All sessions were conducted and recorded via Zoom and later transcribed using OtterAI~\cite{otter}. Upon completing the study, each participant was provided with a compensation of \$25.

\subsection{Measures}
\label{section:study1-measures}
Our analysis focused on comparing LLMs' and humans' performances along the three key outcome measures of CW mentioned in \autoref{section:related-work}. For task completion rates, we calculated the percentage of tasks completed per app. To understand the path navigation pattern differences, we explored two measures. First is the number of steps taken to complete each user task. It reflects the efficiency of completing the task. Second is the Jensen-Shannon Divergence (JS Divergence), which measures the similarity between two distributions~\cite{menendez1997jensen}. It is based on Kullback-Leibler divergence~\cite{kullback1951kullback} and, in this case, it calculates the distance between two navigation pathways. In our study, we used the correct path set as the reference and calculated the JS divergence score of the LLM set and the human set against the correct set. The results illustrate how much LLMs' and humans' path choices differ from the correct paths. The higher the score, the greater the divergence, indicating more differences in the paths. Finally, to understand the number of potential failure points, we qualitatively coded both LLMs' and humans' rationale for taking an action. If the rationale indicates any confusion or uncertainty, we coded it as a potential failure point.

\subsection{Analyses}
To compare task completion, we conducted an ANOVA test on the average completion rate with post-hoc pairwise t-tests, contrasting between the AI conditions and the human conditions. For the number of steps and JS scores, because there was more variability across tasks, we analyzed the data at the task level. We built mixed-effect models, using the number of steps and JS scores as outcome variables, task id and condition-type (human, GPT, or Gemini) as fixed variables, and participant id as a random effect to account for repeated measures.

%To analyze this data, we first captured all screens that the participants and LLM accessed, and documented the order in which they accessed these screens per task. Then, we structured the data in a long format where each row contains the Human ID or AIID, start screen, end screen, the transition action, the step order in the whole process of completing the task, and the explanation for taking this action (pulled from the participant transcripts and LLM outputs). On average, each GPT run experienced 71.3 screens, each Gemini runs experienced 70.7 screens, and each human participant experienced 76.2 screens. And across all LLM and human runs, 70 common screens were commonly visited by each screen. %This shows that our dataset represents similar coverage of screens by both humans and LLM, which establishes the comparison between their performances.  
%To understand their performance differences, we leveraged multiple quantitative measures, as we will explain in \autoref{section:study1-measures}.
% On average, each GPT run experienced 71.33 screens, each Gemini runs experienced 70.66 screens, and each human participant experienced 76.2 screens. Among the LLM set and human set, 70 screens were commonly experienced by all runs. 
% the 5 GPT runs experienced 76 unique screens across the two apps. The Gemini runs experienced 73 unique screens. The humans experienced 85 unique screens. Across all LLM and human runs, there were 70 common screens experienced.

Additionally, to understand why LLMs and humans took the actions they did and to contrast their explanations against each other, the first author conducted thematic analysis on the rationale extracted from both the LLM set and the human set. Initially, the first author extracted key themes from 3 LLM set runs and 5 participant transcripts and discussed the emerging codes with the research team. The team iteratively discussed and refined the codes until deciding on a final codebook~\cite{terry2017thematic}. The first author then used the finalized codebook to analyze all transcripts and all LLM set runs.

\subsection{Results}
\begin{figure*}
    \centering
    \includegraphics[width=400pt]{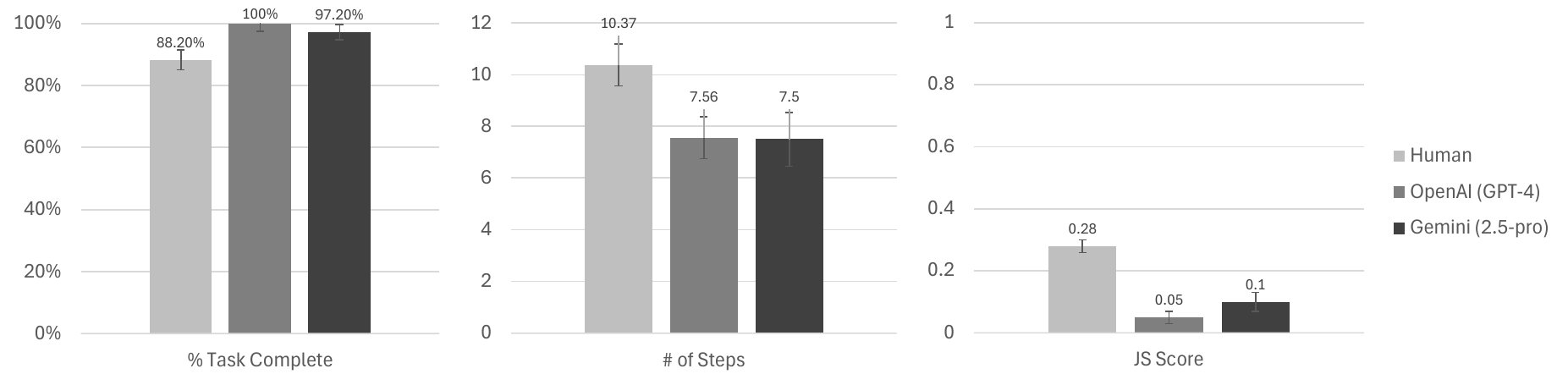}
    \caption{This figure presents the study 1 results, comparing human, GPT, and Gemini runs on task completion rate, number of steps to complete a task, and JS score.}
    \label{fig:completion}
\end{figure*}

As shown in Figure \ref{fig:completion}, regarding completion rate, we found that all but one of the 16 LLM runs was able to navigate the screens and complete all user tasks. Whereas 6 out of 10 human participants encountered some difficulties. On average, humans achieved a lower completion rate ($M = 88.20\%$) than both GPT ($M = 100\%, t = 3.33, p < 0.01$) and Gemini ($M = 97.20\%, t = 2.19, p = 0.04$). We also found that comparing those who did complete the tasks, humans had to take more steps  ($M = 10.37$) than GPT ($M = 7.56, SE = 2.11, t(15.24) = 8.22, p < .001$) and Gemini ($M = 7.50, SE = 2.23, t(13.05) = 7.49, p < .001$) did. When compared to the optimal path, we also found that the the LLMs' walkthroughs were closer to the optimal path than humans, with the humans ($M = 0.28$) achieving a statistically significantly higher JS score than both GPT ($M = 0.05, SE = 0.03, t(13.27) = 7.25, p < .0001$) and Gemini ($M = 0.10, SE = 0.04, t(12.06) = 4.70, p < .001$).

We also found that LLMs noted fewer issues in their think-aloud when navigating the interfaces. On average, human evaluators identified 7.5, whereas GPT runs identified 0.6, and Gemini runs identified 0.5. In total, all human users identified 10 potential failure points, whereas all GPT and Gemini runs aggregated only found 3. This illustrates that LLM achieved a much lower coverage of the potential failure points in contrast to the human users, which again confirms LLMs' ability to navigate the interfaces more successfully and optimally, as it fails to identify the potential breakdowns. For instance, when prompted to \textit{find a lesson about food and drink} on \autoref{fig:app1-explore}, a human participant's (P5's) thought process was: \textit{``the course option or the everyday conversation both seem to be reasonable to contain info about food and drink, though probably the everyday conversation may have more conversational content and the other one may have more targeted words and phrases content related. I'm not sure what to do here.''} P5 explicitly pointed out a potential failure point. However, though LLM similarly observed that the two options -- \textit{tap on course blocks} or \textit{tap on everyday conversation} -- could both serve the purpose here, it did not explicitly mention that this would be confusing. Instead, it was able to prioritize one of the options. LLM shared that \textit{tap on course blocks} is more likely to be the next step than \textit{tap on everyday conversation} because the courses section might offer \textit{``a structured way to locate and select a specific lesson about ordering food and drink.''} %There was no mention of any concern or uncertainty in making these choices, even when it was clear that both options were good candidates that would lead to task completion, and the LLM acknowledged that in its analysis.

However, despite differences in the quantitative measures showing that humans and LLMs navigated the interfaces differently, we did observe that LLMs were capable of providing rationales for their navigation decisions comparable to our human responses. For example, for the task to \textit{change the currency from USD to Euro} in the booking app, P8 described that he would navigate to the \textit{profile} (which is the correct path) because \textit{``that is where I expect where any account setting will be, and that I think for a currency setting, it's usually under account setting.''} Similarly, the LLM also chose to click on \textit{profile} because \textit{``currency settings are typically found in user settings or account preferences, which are usually housed within the profile section of mobile apps.''} This shows that, similar to humans, the LLM is drawing on its knowledge of similar apps. Another example is with the \textit{find a podcast about German} task for the language learning app. On \autoref{fig:app1-explore}, the correct path is to scroll down, and the app would reveal the podcast functionality. P4 explained that she chose to \textit{scroll down} because \textit{``well it's not immediately clear, but looks like there's some more content on this page. Let's scroll down to see.''} LLM provided a similar rationale to \textit{scroll down} as well: \textit{``There may be additional content types (like podcasts) not currently visible on the screen. The illustration implies there's more content below.''} The LLM appeared to be analyzing the problem and making a decision about the next steps, similar to how a human would. 

There were also situations where both LLM and human users were confused about what to do next. When prompted to \textit{change the current course level} in the language learning app (\autoref{fig:app1-home}), P4 expressed concerns that \textit{``I don't see how any one of this would be relevant, maybe it's the profile icon because there might be settings related to that in my account? Though, I think learning plan might also be possible, like that seems to be a course-related feature.''} Similarly, LLM also expressed that \textit{``Users often find account and level settings under their profile icon. This may lead to a settings or preferences page to change the course level. But this is not clearly labeled as related to the course level change. It is possible that other options like explore, review, or learning plan may also have content related to classes and possibly difficulty level.''} Here, LLM is capturing details from the UIs, reporting on challenges with identifying the next step, similar to a human's CW outputs, which illustrates the potential usefulness of LLM's walkthrough think-aloud. 

Given the performance differences between LLM and humans, we further analyzed the qualitative data to explore how their navigation differed and why.

% explain the common screen in the confusion point part

\subsubsection{Human participants were more likely to take a breadth-first search approach when unclear about the possible next actions.}
As discussed, the LLM achieved a higher completion rate than our human participants, fewer steps to complete the user tasks and a lower divergence score from the correct paths. This indicates that the LLM was more efficient than the human participants. One possible reason is that the human participants took breadth-first search approach when they were uncertain about the next steps.

For instance, to perform the task \textit{find a podcast related to German} from the homescreen (\autoref{fig:app1-home}), the correct path is to go to \textit{explore,} where there is a podcast module. The LLM followed this path exactly and provided the following rationale for choosing to interact with the \textit{explore} icon: \textit{```Explore' is likely to lead to a broader set of content types. Podcasts may fall under this category.''} It also explicitly pointed out that \textit{```Explore' is the best candidate for discovering broader content like audio programs or podcasts.''} While it also considered other possibilities such as the \textit{live} icon, it articulated that this option is less likely because \textit{``The `Live' section might include live sessions or events. It's not clear if these are related to podcasts.''} This indicates the LLM was analyzing each potential option in detail and choosing the most likely option, which corresponds to our observation that the LLM's paths were more aligned with those of the correct paths. This could likely be because of how we prompted the LLM as shown in Section~\ref{section:prompt-design}, where we explicitly instructed it to consider all possibilities and provide rationale so that it could complete the CW. 

In contrast, when uncertain about next steps, human users shared a more open and exploratory attitude. In the same setting as the LLM, P4 said \textit{``well, there seems to be multiple things that are relevant here, `live' and `explore' both seem reasonable, I might just try them out first. Or maybe also `review' as well, it might have some review of info on German maybe.''} When further prompted to explain why trying them both out, P4 said, \textit{``as a new user, I just want to be exploratory here, get to know an app, and see what might be the most likely option. So it's not like I have to pick anything, it's more I just want to see what's in which options and explore around.''} In this encounter, P4 expressed a preference to be exploratory and try all options first before committing to a specific path. And P4 did actually first explored the \textit{live} icon in the bottom navigation bar, explaining that \textit{``this could be where all the audio-related things are.''} Then when the information presented did not match her expectation, she said \textit{``then let's try `review' now because it's a broader term and might have the podcasts.''} And finally, she decided to go to \textit{explore} and found the podcast module. When asked why she tried out the options in the order she did, P4 explicitly said that there was no specific ordering and again it is all \textit{``exploratory because I would want to know what's in each option any ways.''}

In another example, on the homescreen (\autoref{fig:app2-home}), the user should move to the search bar to navigate to complete the task ``\textit{find a top-rated nature activity in Amsterdam}'' based on the correct path. P7 did a breadth-first search approach as well when he felt like there are multiple possibilities: \textit{``looks like there's a search bar so I could go there, but also this nature tab on the homescreen seems relevant.''} When asked about the preferences for the next step, P7 said that \textit{``I would just try out everything. It's about being exploratory and knowing what's going on.''} Thus, P7 first attempted to click on the \textit{nature} tab. When no information about Amsterdam was found in that tab, he then used the search bar. Again, during this encounter, the human users' intent was to be more exploratory instead of having to prioritize one option over another. In contrast, the LLM analyzed the situation and decided to choose the \textit{search bar} over the \textit{nature tab} because: \textit{``This action [clicking on the search bar] allows me to directly enter my criteria (`nature related activity in Amsterdam'), which gives the most control over the search results, avoiding the potential detour.''} While it did mention that \textit{``nature activities are likely to be listed under `Nature',''} it also acknowledged that this option is not the most direct option to achieve the task and thus decided to prioritize the other possibility. Therefore, the LLM chose the more obvious or the most likely option.

Overall, when human users encounter confusion points, they may choose to take a breadth-first-search approach, exploring all possible paths instead of having to decide on one option. They did not force themselves to choose one for the next step. This was different from the LLM's more rational behavior of always trying to take the most plausible option. This is likely because the LLM was instructed to make a path selection and specify the next action, whereas human participants did not fully anticipate that they had to make a clear distinction across the different paths. 
\subsubsection{Human memory faults influence overall performance.}
Another potential reason for the LLM's navigation paths being more aligned with the correct path is that humans sometimes misremembered their previous encounters in the apps, leading to a path choice based on faulty recall. For example, for the task \textit{set weekly goal in the language learning app} (starting from \autoref{fig:app1-home}), the correct path is to click on \textit{profile} icon. Yet, P3 decided to click on \textit{learning plan} because she falsely recalled that she saw this functionality in that option when she was performing another task: \textit{``I think I saw it someplace. I don't remember exactly where. I think when we clicked on learning plan?''} After the first failed attempt, P1 then interacted with the \textit{limited offer} button and the \textit{streak} icon before remembering that the functionality is associated with the \textit{profile} icon. This memory issue was the main reason why P1 was unable to follow the correct path to complete the task. Similarly, for the task \textit{``find a lesson about food and drink''} (\autoref{fig:app1-home}) the correct path is to navigate to \textit{learning plan} where there is a relevant course module. P5 initially wanted to click on the \textit{learning plan} since she saw that as more relevant, but then she claimed to remember that she has seen something related to the task from a previous encounter with the \textit{review} icon at the bottom and decided to click on that instead: \textit{``I remember there was this option on the review tab. Let me go and check that.''} Thus, she made a path navigation decision based on her reflection of previous encounters. Yet it was mistaken, and it led her away from the correct path.

In contrast, in both aforementioned cases and across the entire synthetic CW set (LLM set), we did not observe this behavior from the LLM. This is likely because we provided the LLM with the saved history of all the previous steps in a CW. We provided both the screenshots encountered and the LLM's previous decision to move from one screen to another, and the corresponding rationale for that decision, as discussed in \autoref{section:prompt-design}. Thus, it is likely that since we provided context information to the LLM, it was able to ``retain a correct memory'' and therefore behaved differently from the users.

% \subsubsection{LLM identified fewer potential failure points than humans did.}
%\subsubsection{LLM identified fewer confusion points than humans did.}

% \subsubsection{The LLM is less likely to make a wrong selection due to memory issues}
% task level vs. full session level, whereas the LLM did not encounter this issue. This is partially related to how we set up the CW session, where we only provided the LLM one task at a time. 

% P1 said, \textit{``I think I've seen this one here, so let's go to xxx,''} but actually the correct path is nested in the \textit{} component. 
% then chose the wrong action because it had the wrong impression of the user tasks. Moreover, when the participants were completing one specific task, they felt uncertain at times about a screen they visited and wanted to go back to confirm. The LLM did not exhibit this type of behavior since we provided them with the previous screenshots. It understands the context.

% However, the LLM did not exhibit this type of behavior. Specifically, the LLM was very explicit about what it intends to do and why. Given our prompt, the LLM did not get stuck

% Given context window limitations, when we implemented

% P1 

% P3 can we go to review again/

% Give

Overall, we observed differences between how LLM and human users conducted a CW, where LLM achieved a higher task completion rate and navigated more optimal paths. Additionally, we found that the LLM identified fewer potential failure points. Next, we discuss a second study in which we specifically prompted the LLM to identify potential failure points, exploring whether it could predict the human identified potential failure points.

% A potential explanation is that even though LLM was more inclined to choose the most likely correct approach every step of the way, it also considered other possibilities and was able to make a prediction about whether or not that would be confusing for users. 

% synthetic evaluation with expert evaluation, we first contrasted the number of issues identified by synthetic evaluation within the master set to those identified by the expert evaluators. We then analyzed the data by coded heuristic and severity rating in the master set to observe any significant performance differences. In addition to the quantitative analysis, we conducted a series of qualitative analyses on the reported issues across sets to gain further insights into the potential differences between synthetic evaluation and human evaluation.

% similar completion rate but then different paths 

% \section{Study 2: LLM-driven CW can predict human confusion rating}
% In this stage, we tested two main aspects. 

% Given the aforementioned behavior difference, we are interested in further comparing human's and LLM's behavior regarding another key aspect of CW: the confusion points. We define confusion points as the screens on which the users are uncertain/confused about the choice they make. This is a key indicator of a potential usability issue. 

% and 2) whether the user is certain that their action will contribute to completing the task (whether the current screen is a confusion point for them). 

\section{Study 2: Aligning LLM Performance with Human Performance by Introducing a Separate Confusion Rating Mechanism}
We showed in study 1 that LLMs ``walkthrough'' interfaces differently from humans. However, given that LLMs were able to identify possible paths and provide insights about their decisions, we wondered if we can modify our CW design to help identify screens where human users would face navigation challenges, even if LLMs do navigate differently from humans. It may be possible that LLMs, despite being more successful at picking the right paths, can still be used to report on whether some screens are more challenging to navigate. If possible, this would still be a valuable use of LLMs for user researchers in the context of CW. 

Thus, in this study, we sought to explore whether we can prompt LLMs to help identify potential failure points. As an additional exploration, we also explored if we could do so by simply presenting LLMs with one screen at a time, without having it walk through the full set of screens (i.e., without context). If possible, this could be a simpler and cheaper way of using LLMs to gain insights about interface learnability.

\subsection{Updated Prompts with Potential Failure Points Detection}
In this study, we repeatedly ran the with- and without-context versions three times each for both GPT-4 and Gemini-2.5-pro, using the same apps and screens as study 1. Here, we explain the details of the with- and without-context approaches.

\subsubsection{LLM With-Context Set}
We added instructions in our evaluator so that, for each screen as it walks through the interface, it also identifies whether there is a potential failure point by providing a rating for a screen on a three-level scale (not at all confusing, slightly confusing, and very confusing), and the rationale for that rating. Like our evalautor presented in \autoref{section:prompt-design}, when the LLM determines the potentially failure points, it has access to all the interaction history in the CW session, including previous screens visited, all previous transitions, and the rationale for that transition. The updated prompt is shown in \autoref{appendix:with-context prompt} with the changes highlighted.

\subsubsection{LLM Without-Context Set}
For the without-context version, we had a different setup. The evaluator is presented with all of the human-navigated screens from study 1 and the respective tasks (one screen at a time), and is asked to identify potential failure points and provide a rationale. The without-context version is not conducting a walkthrough of the interface, and thus does not have access and is not asked to consider other screens and prior interactions. The prompt for without-context is shown in \autoref{appendix:without-context-prompt}.

\subsection{Measures}
In our prompt, the LLM failure point rating was reported as a three-level scale (not confusing at all, slightly confusing, and very confusing). However, after initial analysis, we found that on average, the LLM only made 2  ``very confusing'' ratings out of the 70 screens it experienced in the GPT runs. Thus, we combined the ``slightly confusing'' and ``very confusing'' levels together, treating it as a binary measure.

\subsection{Analyses}
To test whether LLM's rating was able to predict the potential failure points identified by humans, we built a mixed-effect binomial logistic regression model, using the human potential failure point rating as outcome variable, the LLM potential failure point rating as fixed variable, and task id and participant id together as nested random effect to account for repeated measures. To understand if the with- and without-context LLMs were consistently predicting the potential failure points by humans, we also calculated pairwise Cohen's Kappa scores for all GPT and Gemini runs~\cite{gisev2013interrater}. 

% (\textit{Human Failure Points Rating $\sim$ LLM Failure Points Rating + (1 | TaskID)}) and tested its statistical significance. 

% we tested both the without and with context approaches across repeated prompting thre and across platforms (GPT-4 and Gemini-2.5-pro). Additionally, we calculated the Cohen's Kappa score to understand the inter-rater reliability for the confusion ratings across runs. 

\subsection{Results}

For the with-context approach, all GPT runs and Gemini ones were able to predict human failure points (\autoref{table:effect-size}). For the GPT runs, the odds ratios ranged between 5.45 and 7.5 ($p < 0.01$), meaning that when the LLM identified a potential failure point, the human was 5.45 to 7.5 times more likely to also consider it as a potential failure point. For the Gemini runs, the odds ratios were between 2.22 and 4.72 ($p < 0.05$). Additionally, when examining average pairwise comparisons of Cohen's Kappa inter-rater reliability score, both the GPT runs ($M=0.64, SD=0.06$) and Gemini runs ($M=0.63, SD=0.13$) were moderately consistent in their predictions (\autoref{table:cohen-with}).

%showed that the GPT runs (M=0.65, SD=0.07) and Gemini runs (M=0.65, SD=0.07) achieved a moderate 

%And all LLM runs (both GPT and Gemini runs) statistically significantly predicted the human identified potential failure points (\autoref{table:effect-size}). Additionally, both GPT runs (M=0.65, SD=0.07) and Gemini runs (M=0.65, SD=0.07) achieved a moderate pairwise Cohen's Kappa inter-rater reliability score , indicating consistent performances in predicting human identified potential failiure points.

% And when compared pairwise, participants achieved a moderate Cohen's Kappa inter-rater reliability for OpenAI (M=0.65, SD=0.07) and Gemini (M=0.63, SD=0.13), demonstrating a consistent approach. This highlights that with this additional prompting, our LLM could better align with human performances in finding potential failure points.
\begin{table}[ht]
\centering
\caption{This table provides the effect size of the mixed-effect binomial logistic regression for both with- and without-context LLM runs (both GPT and Gemini runs). 
The * demonstrates the statistical significant level, * represents $p < 0.05$, ** represents $p < 0.01$, *** represents $p < 0.001$. 
If there is no * for an entry, it indicates the model was not statistically significant.}
\begin{tabular}{|l|l|l|l|l|l|l|}
\hline
 & \makecell[l]{GPT\\Run1} 
 & \makecell[l]{GPT\\Run2} 
 & \makecell[l]{GPT\\Run3} 
 & \makecell[l]{Gemini\\Run1} 
 & \makecell[l]{Gemini\\Run2} 
 & \makecell[l]{Gemini\\Run3} \\ 
\hline
With Context 
 & 5.48 *** & 7.50 ** & 5.45 *** & 4.72 *** & 5.48 *** & 2.22 * \\ 
\hline
Without Context 
 & 1.97 * & 1.71 & 1.65 & 1.30 & 1.55 & 2.19 * \\ 
\hline
\end{tabular}
\label{table:effect-size}
\end{table}

For the without-context approach, we observed lower predictability and higher inconsistencies. The log odds for the GPT's predictions were between 1.65 and 1.97, with only the highest run reaching statistical significance in predicting human failure points. The log odds for Gemini's predictions were between 1.30 and 2.19, and again, with only one of the ones significantly predicting human failure points. When examining the average pairwise Cohen's Kappa score (\autoref{table:cohen-without}), we also found that the predictions were more dispersed, for both within GPT runs and within Gemini runs. This points out that the without-context was not only less predictive, but also more inconsistent across runs. 

%When GPT run 1 identified a potential failure point, the human was 1.97 times more likely to also consider it as a potential failure point ($p < 0.05$). But GPT run 2 and run 3 were not able to statistically significantly predict the potential failure points by human. Similarly, for Gemini run 3, it could statistically significantly ($p < 0.05$) predict, whereas Gemini run 1 was not able to do so ($p = 0.37$). Additionally, we observed the pairwise Cohen's Kappa score was highly dispersed, as shown in [CITE] for both within GPT runs (M, SD) and within Gemini runs (M, SD). Moreover, we observed that between GPT and Gemini runs, the agreeability was quite low, with a few receiving a negative inter-rater reliability score. This again points out that the without-context sometimes are able to predict potential confusion points, but sometimes not able to do so.

\begin{table}[ht]
\centering
\caption{Pairwise Cohen's Kappa scores for GPT and Gemini runs.}
\setlength{\tabcolsep}{2pt} % reduce horizontal padding
\renewcommand{\arraystretch}{1.1}

\begin{minipage}{0.4\textwidth}
\centering
\subcaption*{With-context LLM}
\label{table:cohen-with}
\begin{tabular}{|>{\raggedleft\arraybackslash}p{0.9cm}|>{\raggedleft\arraybackslash}p{0.9cm}|>{\raggedleft\arraybackslash}p{0.9cm}|>{\raggedleft\arraybackslash}p{0.9cm}|>{\raggedleft\arraybackslash}p{0.9cm}|>{\raggedleft\arraybackslash}p{0.9cm}|>{\raggedleft\arraybackslash}p{0.9cm}|>{\raggedleft\arraybackslash}p{0.9cm}|}
\hline
 & GPT Run 1 & GPT Run 2 & GPT Run 3 & Gemini Run 1 & Gemini Run 2 \\
\hline
GPT Run 2 & \cellcolor{gray!20}0.65 & & & & \\
\hline
GPT Run 3 & \cellcolor{gray!20}0.64 & \cellcolor{gray!20}0.53 & & & \\
\hline
Gemini Run 1 & \cellcolor{gray!60}0.44 & \cellcolor{gray!60}0.40 & \cellcolor{gray!60}0.43 & & \\
\hline
Gemini Run 2 & \cellcolor{gray!60}0.38 & \cellcolor{gray!60}0.51 & \cellcolor{gray!60}0.43 & \cellcolor{gray!20}0.51 & \\
\hline
Gemini Run 3 & \cellcolor{gray!60}0.45 & \cellcolor{gray!60}0.50 & \cellcolor{gray!60}0.56 & \cellcolor{gray!20}0.77 & \cellcolor{gray!20}0.62 \\
\hline
\end{tabular}
\end{minipage}
\hfill
\begin{minipage}{0.5\textwidth}
\centering
\subcaption*{Without-context LLM}
\label{table:cohen-without}
\begin{tabular}{|>{\raggedleft\arraybackslash}p{0.9cm}|>{\raggedleft\arraybackslash}p{0.9cm}|>{\raggedleft\arraybackslash}p{0.9cm}|>{\raggedleft\arraybackslash}p{0.9cm}|>{\raggedleft\arraybackslash}p{0.9cm}|>{\raggedleft\arraybackslash}p{0.9cm}|>{\raggedleft\arraybackslash}p{0.9cm}|>{\raggedleft\arraybackslash}p{0.9cm}|}
\hline
 & GPT Run 1 & GPT Run 2 & GPT Run 3 & Gemini Run 1 & Gemini Run 2 \\
\hline
GPT Run 2 & \cellcolor{gray!20}0.62 & & & & \\
\hline
GPT Run 3 & \cellcolor{gray!20}{-0.09} & \cellcolor{gray!20}{-0.02} & & & \\
\hline
Gemini Run 1 & \cellcolor{gray!60}0.26 & \cellcolor{gray!60}0.10 & \cellcolor{gray!60}{-0.04} & & \\
\hline
Gemini Run 2 & \cellcolor{gray!60}0.28 & \cellcolor{gray!60}0.11 & \cellcolor{gray!60}{-0.03} & \cellcolor{gray!20}0.39 & \\
\hline
Gemini Run 3 & \cellcolor{gray!60}0.10 & \cellcolor{gray!60}{-0.05} & \cellcolor{gray!60}{-0.04} & \cellcolor{gray!20}0.38 & \cellcolor{gray!20}0.24 \\
\hline
\end{tabular}
\end{minipage}
\end{table}

\subsubsection{With-context LLM more consistently predicted the potential failure points.}
As discussed, LLM was making a more inconsistent prediction when it did not have enough context about the prior screens it had visited in a CW session. Without context, it was difficult for the LLM to infer its current state and decide its next action. For instance, when asked to \textit{find an outdoor day trip in Amsterdam and save it to view later}, on the search screen \autoref{fig:app2-search}, the without-context LLM identified this as a failure point and reported that: \textit{``This screen appears to be a location selection page, which is a necessary step in narrowing results to activities in a specific city. I do see Amsterdam listed, so I could tap on it and expect to go to the next screen where I can browse activities. However, at this point, there's no indication that selecting Amsterdam will lead me to filters like `outdoor' or `day trip,' or options to save a result for later. \ldots While this screen is logically part of the task flow, the lack of visible guidance toward the full goal creates mild uncertainty.''} This confusion seems to stem from a lack of knowledge of its subtask, what it needs to be doing specifically on a screen, relative to the task, and an assumption that it should be conducting filtering even though it is on a search screen. In contrast, the with-context LLM demonstrated a clear understanding of where it is in conducting the task and thus did not identify this as a potential failure point: \textit{``The purpose of this screen is clear — I can either search manually or select a city from the list.''} This contrast highlights the context information's impact on the LLM's performance, and thus explains why the with-context LLM predicted potential failure points more consistently than the without-context LLM.

\section{Discussion}
Recently, there has been growing attention towards the use of LLM to scale usability testing. While existing works have tested the competency of LLM in navigating through mobile apps~\cite{eskonen2020automating, zhong2025synthetic, guerino2025can, duan2024generating}, there remains an open question of whether LLM could simulate human behavior in performing usability testing such as CW. Our work provides empirical data showing that LLM does not fully simulate human behavior -- and the differences are meaningful for usability evaluation purposes. In fact, LLM outperformed humans in both task completion rate and path navigation patterns, and reported fewer issues with navigation. This shows that the LLM excelled at navigating apps, but not so much in simulating human behavior in conducting CW. Traditionally, people have designed and trained LLMs to optimize for the best performances (e.g., high task completion rate, optimal path navigation, etc.)~\cite{ran2025beyond, liu2024make, abukadah2024mapping, zhang2025appagent}. However, in usability testing, especially for CW where the purpose is to understand how to best design UIs, it is more important that the LLM be able to simulate human behavior. It is not about perfect performances, but rather about identifying usability breakdowns regarding learnability of an interface. As we have shown, the LLM is still lacking from this perspective.

% In contrast to HE that evaluates static screens, CW focuses more on the learnability of an interface. Thus, CW results heavily rely on 

% Another possibility of explaining this 

% impltested a ``naive'' version of our prompt where we provided the LLM with a minimum set of instructions. It only specifies to conduct a CW, without instructing the LLM about what to consider and what format the output should be in. Our results revealed that, with this prompting, the task completion rate was still 100\%, but the JS divergence score against the correct path was higher than our LLM at xxx. Yet regardless, its JS divergence score against humans is still high at xxx. Upon qualitative analysis, we observed that the LLM made a few random choices in its navigation without concrete reasoning. This is potentially because in this ``naive prompt,'' we did not explicitly instruct it to consider all possibilities. Therefore, our prompt was effective in improving LLM's performance in conducting CW and that our prompting did not limit LLM's performance such that it became unable to simulate human behavior. More work is still needed to uncover the root cause of why the LLM was unable to simulate human behavior. 

% - prompting can make LLMs behave more rationally -> optimize path completion but undermines human-likeliness

There were two qualitative insights that illustrated the differences between LLM and humans. The first one is humans took a more breadth-first search approach when uncertain about next steps, whereas LLM acted more rationally, only choosing the most likely option. Usually, as people navigate an app, they would analyze different signals from different components on screen, and make a decision based on which signal they believe is most likely to lead to completing the task. When the UIs are clear and no potential failure points are present, as we showed in our results, both LLMs and humans are able to behave rationally and pick out the most possible path. However, when the UIs have multiple competing components with similar probabilities to lead to task completion (and thus potential failure points), we observed differences. Humans presented a more exploratory behavior, explaining that they would want to know what each path actually contains before making a decision. LLM maintained its completely rational behavior, always choosing one of the more probable options, despite the fact that when it analyzed the screens, it also acknowledged that there were multiple options with similar probability leading to task completion. This observation raises the question of why LLM is performing so rationally.

One possibility is that this was due to the prompting we did. Recall that in our prompt (\autoref{section:prompt-design}), to address the issue that the LLM behaved randomly with a naive prompt, we explicitly instructed the LLM to consider all possible next steps, and explain why each possibility is relevant to the task before making a path selection. It is possible that while this prompt improved LLM's performances in navigating the apps, it also made the LLM behave more rationally (i.e., always pick the next step with the highest likelihood to help it achieve the task) and thus less likely to explore the interface and simulate human behavior. This raises the question of whether this is inherent to how LLM works or if it is related to how we designed our prompt. The important question is whether and how LLM could be prompted to simulate human behavior. Given that LLM is trained on human data, it might be possible to prompt it to ``make mistakes'' or ``be more exploratory'' like humans do. One potential exploration could be finetuning the LLM with more targeted samples about how humans behave less rationally using a one-shot or few-shot approach ~\cite{dang2022prompt}. 

Another observation we made was that the humans sometimes misremembered where a feature was nested and/or misremembered information on a specific screen, which led them to explore the wrong path or repeatedly move back and forth on the same path. The LLM did not experience this issue because we provided it with all the history information within a CW session (including prior screens visited). This context is very important because it essentially serves as an accurate mental map of the app, representing all the screens explored. If the human users had access to all the previous screens visited and all the previous actions performed, it is highly likely that they may not have experienced memory issues either. But this does raise an interesting question of whether LLM should be designed to actively simulate this type of memorization behavior in CW. Future work could design prompts and/or mechanisms to simulate this behavior, and test how designers respond to this type of LLM output, leveraging it to quickly iterate on UI design. 

% especially regarding whether they find the output can potentially identify failure points in UI design. This will also address the question of how much LLM could really ``think'' like a human being, given all the uncertainties and sometimes randomness with humans' behavior (e.g., memory issue, etc.)

Despite these performance differences, our work did reveal that when prompted specifically during CW, LLM can be used to help predict where the human identified potential failure points are within the UIs. We also found that this held true consistently for the with-context version and not as consistently for the without-context version. There are two implications for this. First, though LLM may not be fully simulating human behavior in completing the tasks and navigating the apps, designers could still use it to identify potential failure points, which is an important measure of CW outcome. Thus, in some sense, LLM-supported CW is still useful. The key is how designers analyze the LLM output to inform design changes. Second, the with-context LLM performed more consistently than the without-context LLM. This indicates that for the LLM to perform CW effectively, keeping track of the navigation history (including previous screens visited, transition between screens) is important. It helps LLM assess what subtask it is performing at any given time, which then helps LLM better assess the learnability of an interface. 

\section{Limitations}
In our studies, as mentioned in \autoref{section:prompt-design}, for the apps tested, we manually set up a dataset containing all the screens and transitions across screens to help facilitate the walkthrough. While our approach is well-suited for early prototypes, with recent tools like the Claude computer use \cite{claude-computer-use} and LLM supported navigation tools~\cite{ran2025beyond, ye2025mobile, browser-use, webagent}, there is a possibility to build out an end-to-end system that automatically captures screens as a separate navigator agent performs the navigation action. Future work could build this tool and leverage it to support fully automated CW. 

Additionally, we only tested our approach on mobile interfaces. LLMs may exhibit different behavior with web interfaces, given the design differences across mobile and web interfaces. Future research could test the adaptability of our approach to web UIs and further explore behavioral differences between human and LLM.
\section{Conclusion}
Our study conducted a comparative study, contrasting LLMs' and human's behavior in conducting CW. We found that while LLM does not simulate human's behavor in doing CW (as it outperforms humans in path completion rate and path navigation), it could be additionally prompted to predict the potential failure points identified by humans. This indicates that LLM could still be leveraged for scaling usability testing such as CW, as long as designers are aware of how they behave differently and use it only for where it could align with humans. 
% Conducting usability testing like cognitive walkthrough (CW) can be costly. Recent developments in large language models (LLMs), with visual reasoning and UI navigation capabilities, present opportunities to automate CW. We explored whether LLMs (GPT-4 and Gemini-2.5-pro) can simulate human behavior in CW by comparing their walkthroughs with human participants. While LLMs could navigate interfaces and provide reasonable rationales, their behavior differed from humans. LLM-prompted CW achieved higher task completion rates than humans and followed more optimal navigation paths, while identifying fewer potential failure points. However, follow-up studies demonstrated that with additional prompting, LLMs can predict human-identified failure points, aligning their performance with human participants. Our work highlights that while LLMs may not replicate human behaviors exactly, they can be leveraged for scaling usability walkthroughs and providing UI insights, offering a valuable complement to traditional usability testing.
\bibliographystyle{ACM-Reference-Format}
\bibliography{10-bibliography}
\appendix
\section{LLM Prompt}

\label{appendix}

% \textbf{image selector}
% \begin{quote}
% \begin{displayquote}
%     \texttt{
%     You are helping match a user's described action with one of the app's transition options. The user said they would \{next\_action\_text\}. Available transitions are: \{list of possible next steps\}. Return the best matching transition as plain text. If none are relevant, return ``NONE''
%     }
% \end{displayquote}
% \end{quote}
\subsection{Pipeline Component Prompts}
\subsubsection{Facilitator Prompt}
\label{appendix:failitator-prompt}
\begin{displayquote}
    \texttt{You are a facilitator for a cognitive walkthrough. You will provide the users with a screenshot of a mobile application and a text description of the user task: [Task description]. You will ask the users to discuss which component they will interact with next and why they would like to do so to complete the given user task. If the user discusses more than one possible action, ask them to provide a rating of how confident they believe the next action will be to complete the task, on a scale of low, medium, and high. Tell them they may be exploratory and unsure. If the user cannot decide what to do next, do not analyze or provide your own opinion at any point in the simulation. Simply ask them to rethink. Remain neutral and prompt them to think further—ask them why and how questions. If the users’ responses lack detail or the rationale does not make sense, prompt them for refinement. You can remind the users of the ``think aloud'' protocol. Make sure the users only discuss the next possible action, while thinking through the general direction of their actions. If the users repeatedly provide the same action, remind them they are repeating and that they can scroll and/or exit the current screens to go back to the main screen. If they are still stuck in a specific place, prompt them to rethink their actions. Ask the user whether they have completed the task. If so, terminate the walkthrough. Remind the user to provide reasoning in JSON format. Do not provide example outputs, and do not remind them of the current screen information—keep your instruction simple. As the facilitator, do not describe the current screen at all. If the user has already attempted a task with no effect but tries to go down the same path again, stop them from doing so and ask them to rethink their actions. If the user is stuck on the same screen, ask them to consider whether a specific icon or button is already activated.
 }
\end{displayquote}

\subsubsection{Evaluator Prompt}
\label{appendix:evaluator-prompt}
\begin{quote}
\begin{displayquote}
    \texttt{
    You are helping with a cognitive walkthrough. You are given a screenshot of a mobile app and a user task: [Task description]. Your job is to explain what you would do next, considering options you see. Describe if the current screen is confusing in any way that may prevent you from achieving the task, provide rationale. Use the think aloud protocol. Be exploratory with your choices. You must respond in the following JSON format: \{current\_state: \textless{}string\textgreater{}, possible\_actions: \{action: \textless{}string\textgreater{}, rationale: \textless{}string\textgreater{}, confidence:  \textless{}low|medium|high\textgreater{}, next\_action: \textless{}string\textgreater{}, next\_action\_rationle: \textless{}string\textgreater{}, \} If you see the same screen as before, it means that the action did not have any impact on the app. Consider trying other actions. If you plan to go back to a previous screen (for example, by clicking a tab or back arrow), clearly indicate this in next\_action. Likewise, if you plan to stay on the same screen, describe that explicitly too. If you are stuck on one screen, you can always exit and find other paths— especially by going back to the main screen or scrolling down to see more options. If one type of interaction doesn’t work out, consider trying other interactions such as clicking, swiping, or selecting another possible element. If you see the same screen as before, it means that the action did not have any impact on the app. Consider trying other actions. If you plan to go back to a previous screen (e.g., by clicking a tab or back arrow), clearly indicate this in next\_action. Note that if you are stuck on one screen, you can always exit and find other paths, especially by going back to main, or scroll down to see more options. If one type of interaction doesn't work out, consider trying other interactions such as clicking, swiping, etc., or click on another possible element.
    }
\end{displayquote}
\end{quote}

\subsection{Study 2: With- and Without- Context LLM Prompts}
\subsubsection{With-context Prompt}
\label{appendix:with-context prompt}
\begin{quote}
\begin{displayquote}
    \texttt{
    You are helping with a cognitive walkthrough. You are given a screenshot of a mobile app and a user task [Task description]. Your job is to explain what you would do next, considering options you see. Use the think aloud protocol. Be exploratory with your choices. You must respond in the following JSON format: Describe if the current screen is confusing in any way that may prevent you from achieving the task, provide rationale. \hl{Provide the confusion rating on the following scale: not at all confusing, slightly confusing, very confusing.} You must respond in the following JSON format: {current state: <string>; possible actions: {action: <string>,rationale: <string>,confidence: <low|medium|high>}; next action: <string>; next action rationale: <string>; \hl{confusing or not: <string>; confusing or not rationale: <string>}} \hl{For the confusing or not rationale, more than just the layout of the screen, think about how this screen may or may not contribute to the final outcome, whether it is anticipated that this screen will lead to other reasonable screens, whether it is within the process of accomplishing a task.} If you see the same screen as before, it means that the action did not have any impact on the app. Consider trying other actions. If you plan to go back to a previous screen (for example, by clicking a tab or back arrow), clearly indicate this in next\_action. Likewise, if you plan to stay on the same screen, describe that explicitly too. If you are stuck on one screen, you can always exit and find other paths— especially by going back to the main screen or scrolling down to see more options. If one type of interaction doesn’t work out, consider trying other interactions such as clicking, swiping, or selecting another possible element. If you see the same screen as before, it means that the action did not have any impact on the app. Consider trying other actions. If you plan to go back to a previous screen (e.g., by clicking a tab or back arrow), clearly indicate this in next\_action. Note that if you are stuck on one screen, you can always exit and find other paths, especially by going back to main, or scroll down to see more options. If one type of interaction doesn't work out, consider trying other interactions such as clicking, swiping, etc., or click on another possible element.
    }
\end{displayquote}
\end{quote}

\subsubsection{Without-Context Prompt}
\label{appendix:without-context-prompt}
\begin{quote}
\begin{displayquote}
    \texttt{You are helping with a cognitive walkthrough. You are given a screenshot of a mobile app and a user task [Task description]. Your job is to explain based on the user task and the current screen you see, describe if the current screen is confusing in any way that may prevent you from achieving the task, provide rationale. Provide the confusion rating on the following scale: not at all confusing, slightly confusing, very confusing. Use the `think aloud' protocol. You have to provide a rationale for why you provide such a rating. You must respond in the following JSON format:{confusing or not: <string>, confusing or not rationale: <string>}} \texttt{For the confusing or not rationale, more than just the layout of the screen, think about how this screen may or may not contribute to the final outcome, whether it is anticipated that this screen will lead to other reasonable screens, whether it is within the process of accomplishing a task.}
\end{displayquote}
\end{quote}

\end{document}